# Defining Reference Sequences for *Nocardia* Species by Similarity and Clustering Analyses of 16S rRNA Gene Sequence Data

**Manal Helal[1,2], Fanrong Kong[2], Sharon C. A. Chen[1,2], Michael Bain[3], Richard Christen[4], Vitali Sintchenko[1,2]***

1 Sydney Medical School, The University of Sydney, Sydney, New South Wales, Australia, 2 Centre for Infectious Diseases and Microbiology, Westmead Hospital, Sydney West Area Health Service, Sydney, New South Wales, Australia, 3 School of Computer Science and Engineering, University of New South Wales, Sydney, New South Wales, Australia, 4 University of Nice Sophia-Antipolis, and CNRS UMR6543, Parc Valrose, Centre de Biochimie, Nice, France

## Abstract

*Background:* The intra- and inter-species genetic diversity of bacteria and the absence of 'reference', or the most representative, sequences of individual species present a significant challenge for sequence-based identification. The aims of this study were to determine the utility, and compare the performance of several clustering and classification algorithms to identify the species of 364 sequences of 16S rRNA gene with a defined species in GenBank, and 110 sequences of 16S rRNA gene with no defined species, all within the genus *Nocardia*.

*Methods:* A total of 364 16S rRNA gene sequences of *Nocardia* species were studied. In addition, 110 16S rRNA gene sequences assigned only to the *Nocardia* genus level at the time of submission to GenBank were used for machine learning classification experiments. Different clustering algorithms were compared with a novel algorithm or the linear mapping (LM) of the distance matrix. Principal Components Analysis was used for the dimensionality reduction and visualization.

*Results:* The LM algorithm achieved the highest performance and classified the set of 364 16S rRNA sequences into 80 clusters, the majority of which (83.52%) corresponded with the original species. The most representative 16S rRNA sequences for individual *Nocardia* species have been identified as 'centroids' in respective clusters from which the distances to all other sequences were minimized; 110 16S rRNA gene sequences with identifications recorded only at the genus level were classified using machine learning methods. Simple kNN machine learning demonstrated the highest performance and classified *Nocardia* species sequences with an accuracy of 92.7% and a mean frequency of 0.578.

*Conclusion:* The identification of centroids of 16S rRNA gene sequence clusters using novel distance matrix clustering enables the identification of the most representative sequences for each individual species of *Nocardia* and allows the quantitation of inter- and intra-species variability.

Funding: This research was supported by the Australian National Health and Medical Research Council. The funders had no role in study design, data collection and analysis, decision to publish, or preparation of the manuscript.

Competing Interests: The authors have declared that no competing interests exist.

* E-mail: vitali.sintchenko@swahs.health.nsw.gov.au

## Introduction

Sequence-based identification of bacteria relies on the postulate that two strains with matching sequences of house-keeping genes, such as the 16S rRNA gene, are likely to belong to the same species of bacteria. However, if their sequences are different, then the interpretation depends on the magnitude of the difference: minor variations may still represent strains that are closely related taxonomically but more significant differences suggest that the strains might belong to evolutionary related but distinct species of bacteria. The quality of sequence-based identification depends on the representativeness of sequence libraries and the reliability of species definitions, both of which continue to pose challenges [1,2,3]. To improve sequence-based species identification, there has been strong impetus on developing standardised sequences, including nucleotide polymorphisms for a particular species. This process requires the establishment of "reference sequences" or "DNA barcodes" for species identification and recognition of intra-species sequence polymorphisms or "sequence types" [1,2]. Evidence suggests that such a process of curation, in which the designated most representative sequence of a species (or the "centroid" sequence) is derived from discrete "species groups" of sequences, can be automated (e.g., Integrated Database Network System SmartGene), and improves the species-level identification of clinically relevant *Nocardia* isolates [3]. Both major sources of genotypic variation – vertical and horizontal gene tranfer [4] – should be considered when the target sequences for sequence-based identification are selected.

A large number of sequences from different microbial species have been accumulated in public databases. The task of their classification in a diagnostic laboratory has been complicated by the variability in sequences within a species and the absence of 'reference sequences' or the most representative sequences for that species [5,6]. To this end, significant advances have been made in gene annotation and similarity/dissimilarity assessments [7,8] and in the use of different distance metrics [8–10]. Multiple sequence alignment (MSA) methods [11–14] have been proposed as clustering methods, since they arrange sequences according to similarity.

Bacteria of the genus *Nocardia* cause a range of infectious diseases including localised lung and skin infections and disseminated disease [15]. The accurate identification of clinical isolates is critical for diagnosis, prediction of antimicrobial susceptibility and for epidemiological tracking of isolates. Since standard phenotypic identification methods are time-consuming and imprecise [16], nucleic acid-amplification tools targeting conserved gene regions have been developed to facilitate accurate species determination. Of these, 16S rDNA sequence analysis is the most frequently used method for the definitive species identification of *Nocardiae* [16,17]. These methods have led to substantial species re-assignment within the genus, especially among "*Nocardia asteroides*" and "*Nocardia transvalensis*" isolates. For example, two distinct new species, *Nocardia wallacei* and *Nocardia blacklockiae*, have recently been proposed to replace previous *N. asteroides* drug pattern IV and *N. transvalensis* new taxon 1, respectively [18]. Distinct species formerly classified within "*N. asteroides*" include *Nocardia cyriacigeorgica* and *Nocardia abscessus* [16]. Over 80 species have now been described, (http://www.ncbi.nlm.nih.gov/Taxonomy/; http://www.bacterio.cict.fr/n/nocardia.html) [16]. Although numerous *Nocardia* 16S rDNA sequences have been deposited in public gene databases such as DDBJ/EMBL/GenBank consortium, a substantial proportion of bacterial 16S rRNA gene sequences represents inaccurate entries [19,20]. Further, sequence-based analyses are complicated by the lack of consensus regarding the degree of sequence similarity required for the species definition of *Nocardia* [7,21].

The aim of this study was to numerically redefine the species memberships and to identify the most representative 16S rRNA gene sequences for each *Nocardia* species and to compare the performance of clustering algorithms in this classification task. We also aimed to assign *Nocardia* 16S rRNA gene sequences, deposited in the DDBJ/EMBL/GenBank consortium but not fully identified at the time of submission, to individual species using our classification algorithms.

## Results

### Interspecies sequence similarity among Nocardia

Assuming that the assignment of sequences to species is correct and the successful numerical clustering needs to match the species identity as determined by GenBank-comparisons as much as possible, the accepted clustering results of the 364 sequences of 16S rRNA genes produced 80 clusters (for the full list of clusters, refer to Table S1). This was produced by the linear mapping of the alignment distance matrix (LM) algorithm. The majority of the clusters were relatively small in size: only 23 clusters contained five or more sequences of 16S rRNA genes and only two clusters (*N. cyriacigeorgica* and *N. farcinica*) had more than 20 sequences (Figure 1). The LM algorithm had two sensitivity parameters: the number of indices to map to, and the number of indices to include in each cluster. The accepted 80 clusters were achieved when the first parameter was 128 indices, with only one index per cluster. Different sensitivity parameters produced a wide range of potential clusters (Table S2). Table 1 specifies the top twenty seven clusters at the accepted sensitivity level, their sizes and their reference sequence species and corresponding GenBank accession numbers assigned to those sequences. It also shows the proportion of the misclassified sequences, i.e. those whose identification (at the time of initial submission to GenBank) did not match the predominant *Nocardia* species in the cluster. Some of 16S rRNA gene sequences of seventeen *Nocardia* species were co-clustered with other species (Table 2). The high-level 'heatmap' of the distance matrix indicating rectangles around all of the clustering parameters is shown in the electronic supplements (Figure S1).

The alignment of 16S rRNA gene sequences identified hyper-variable regions surrounding V2, V4 and V6 that corresponded to the first 250 base pairs of the 16S rRNA gene, and about 500 and 650-bp, respectively (Figure 2). When principal component analysis (PCA) was applied to separate individual species by these discriminatory structures in the 16S rRNA gene, the first principal component represented 97.11% of the structure in the distance matrix. The second and the third components represented 1.61% and 0.44% of the data structure, respectively. The two-dimensional plot of the 1st and 2nd PCA (as shown in Figure 3-A to visually identify the clusters) was not informative. However, the combination of the 1st and 2nd PCA with the 2nd and 3rd PCA allowed the allocation of sequences into 80 clusters (Figure 3-B). When the 1st and 2nd PCA scores were used as the dataset for linear mapping clustering with a Euclidian distance measure, 67 clusters were identified. K-means method produced the highest accuracy when $k$ was between 27–36 clusters, but we used 77 distinct clusters to match the known GenBank submitted species names.

The relationships between *Nocardia* species clusters were also visualised using a "mountain view" (Figure 4), where each cluster was represented as a peak in the three dimensional terrain. The Cluto algorithm identified ten 'mountain peaks' for the optimal 10 clusters. The clusters with the highest sequence similarity were ones consisting of *N. asteroides* complex sequences (Figure 4; red peaks such as AF430025, AF430026, X84850, X80606, Z36934, AF430019, and DQ659898), and the other neighbouring clusters. Other *Nocardia* species were represented as less homogenous clusters with more diverse sequences of the 16S rRNA gene. Such representation illustrates the relative distance between species highlighting, for example, the uniqueness of two sequences of *N. seriolae* (Accession Numbers AB060281 and AB060282) which appeared as outliers to the rest of the data set (the PCA plots in Figure 3 and the mountain view in Figure 4).

### Performance of clustering algorithms

The linear mapping (LM) achieved the highest performance to 128 indices, and only one index per cluster. This algorithm classified the set of 16S rDNA sequences into 80 clusters, the majority of which corresponded well with the original GenBank species; 304 sequences out of the 364 sequences (83.52%) were clustered with sequences sharing the same species name. When partial matching was taken into account (clusters containing equal numbers of sequences of originally different species names), the classification performance improved further to 93.13%. The LM algorithm outperformed both Cluto and hierarchical clustering over 77 clusters (Table 3 and Table S4). The k-means algorithm with k = 77 produced eight clusters with negative silhouette values (a measure of cluster coherency), indicating the presence of misclassified outliers, while the majority of clusters demonstrated silhouette values higher than 0.4, suggesting clusters sufficiently separated from their neighbours (Figure S2). The average

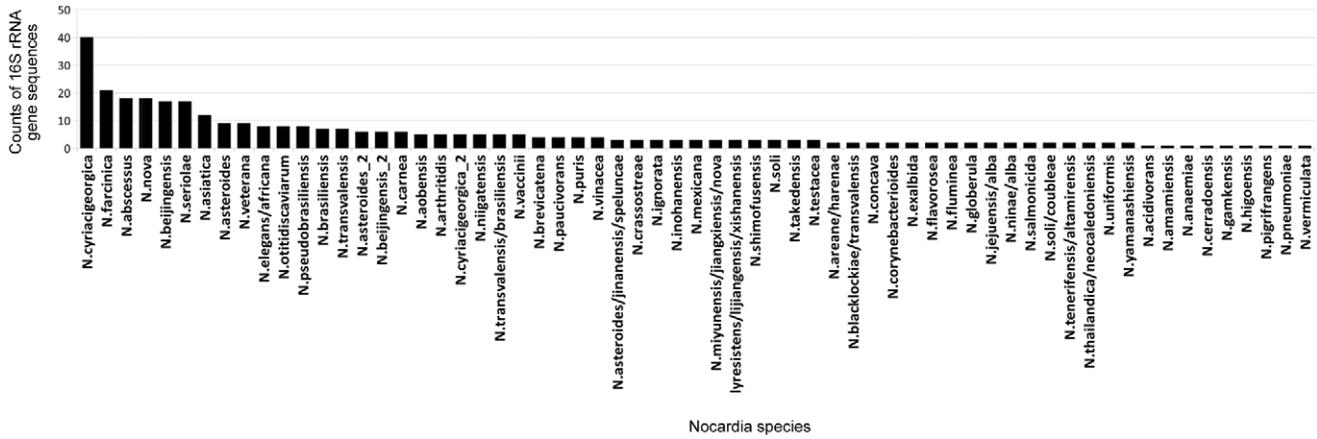

**Figure 1. Clusters of different *Nocardia* species determined by MSA comparisons.**
doi:10.1371/journal.pone.0019517.g001

**Table 1.** *Nocardia* species clusters and their reference sequences*.

| Cluster name | Number of sequences in the cluster* | Reference Sequence Name | GenBank Accession Number | Proportion of other *Nocardia* species assigned to the cluster, % |
|---|---|---|---|---|
| *N. cyriacigeorgica* Cluster 1 | 40 | *N. cyriacigeorgica* | AJ508414 | 5.00 |
| *N. farcinica* | 21 | *N. farcinica* | X80595 | 9.52 |
| *N. abscessus* | 18 | *N. abscessus* | AB212947 | 11.11 |
| *N. nova* | 18 | *N. nova* | X80593 | 0.0 |
| *N. beijingensis* Cluster 1 | 17 | *N. beijingensis* | DQ659901 | 0.0 |
| *N. seriolae* | 17 | *N. seriolae* | X80592 | 0.0 |
| *N. asiatica* | 12 | *N. asiatica* | DQ659897 | 0.0 |
| *N. wallacei* | 9 | *N. wallacei* | Z82229 | 44.44 |
| *N. veterana* | 9 | *N. veterana* | AF278572 | 22.22 |
| *N. elegans* | 8 | *N. elegans* | DQ659905 | 50.0 |
| *N. otitidiscaviarum* | 8 | *N. otitidiscaviarum* | X80599 | 0.0 |
| *N. pseudobrasiliensis* | 8 | *N. pseudobrasiliensis* | DQ659914 | 0.0 |
| *N. brasiliensis* | 7 | *N. brasiliensis* | X80591 | 14.29 |
| *N. transvalensis* Cluster 1 | 7 | *N. transvalensis* | X80598 | 0.00 |
| *N. carnea* | 6 | *N. carnea* | AF430035 | 0.00 |
| *N. asteroides* | 6 | *N. asteroides* | DQ659898 | 16.67 |
| *N. beijingensis* Cluster 2 | 6 | *N. asteroides**** | Z82228 | 16.67 |
| *N. cyriacigeorgica* Cluster 2 | 5 | *N. cyriacigeorgica* | Z82218 | 40.0 |
| *N. arthritidis* | 5 | *N. arthritidis* | DQ659896 | 40.0 |
| *N. vaccinii* | 5 | *N. vaccinii* | AF430045 | 0.0 |
| *N. aobensis* | 5 | *N. aobensis* | AB126876 | 0.0 |
| *N. niigatensis* | 5 | *N. niigatensis* | AB092563 | 0.00 |
| *N. transvalensis* Cluster 2 | 5 | *N. otitidiscaviarum**** | AB201303 | 60.0 |
| *N. brevicatena* | 4 | *N. brevicatena* | AF430040 | 0.0 |
| *N. paucivorans* | 4 | *N. paucivorans* | AF430041 | 0.0 |
| *N. puris* | 4 | *N. puris* | AB097455 | 0.0 |
| *N. vinacea* | 4 | *N. vinacea* | AB162802 | 25.0 |

Note: Rows of *Nocardia* species assigned to more than 1 cluster are highlighted in grey.
*Only clusters with more than three 16S rRNA gene sequences are shown. For the full list refer to Table S3.
**One *N. otitidiscaviarum* 16S rRNA gene sequence in a cluster of five, suggestive of misclassification in the original submission.
***One *N. asteroides* 16S rRNA gene sequence in a cluster of six, other five sequences belong to *N. beijingensis*. Suggestive of misclassification in the original submission.
doi:10.1371/journal.pone.0019517.t001

**Table 2.** Species names of 16S rRNA gene sequences that were co-clustered with other species.

| Nocardia Species Names and GenBank Accession Number(s) | Co-Clustered with Nocardia species |
|---|---|
| N. caishijiensis (AF459443) | N. mexicana (AY555577 and AY560655) |
| N. sienata (AB121770) | N. testacea (AB121769 and AB192415) |
| N. pseudovaccinii (AF430046) | N. vinaceae (AB162802 and AB024312) |
| N. jinanensis (DQ462650) | N. asteroides (Z82231) and N. speluncae (AM422449) |
| N. lijiangensis (AY779043) | N. xishanensis (AY333115) and N. polyresistens (AY626158) |
| N. araoensis (AB108779) | N. arthritidis (AB212949 and AB108781) |
| N. alba (EU249584) | N. jejuensis (AY964666) |
| N. alba (AY222321) | N. ninae (DQ235687) |
| N. jiangxiensis (AY639902, DQ840027) | N. nova (DQ840028 and DQ840029) |
| N. coubleaea (DQ235688) | N. soli (AF277223) |
| N. cummidelens (AF277202) | N. soli (AF277199 and AF430051) |
| N. altamirensis (EU006090) | N. tenerifensis (AJ556157) |
| N. iowensis (DQ925490) | N. brasiliensis (X80591, AY245543 and AF430038) |
| N. neocaledoniensis (AY282603) | N. thailandica (AB126874) |
| N. wallacei (EU099357) | N. transvalensis (AB 084444, AB08445 and AB084446) |
| N. kruczakiae (AY441974, DQ659909) | N. veterana (AF278572, DQ659918 and AY191253) |
| N. africana (AF277198, AF430054, AF302232, AY089701) | N. elegans (AJ854057, DQ659905 and AB237142) |

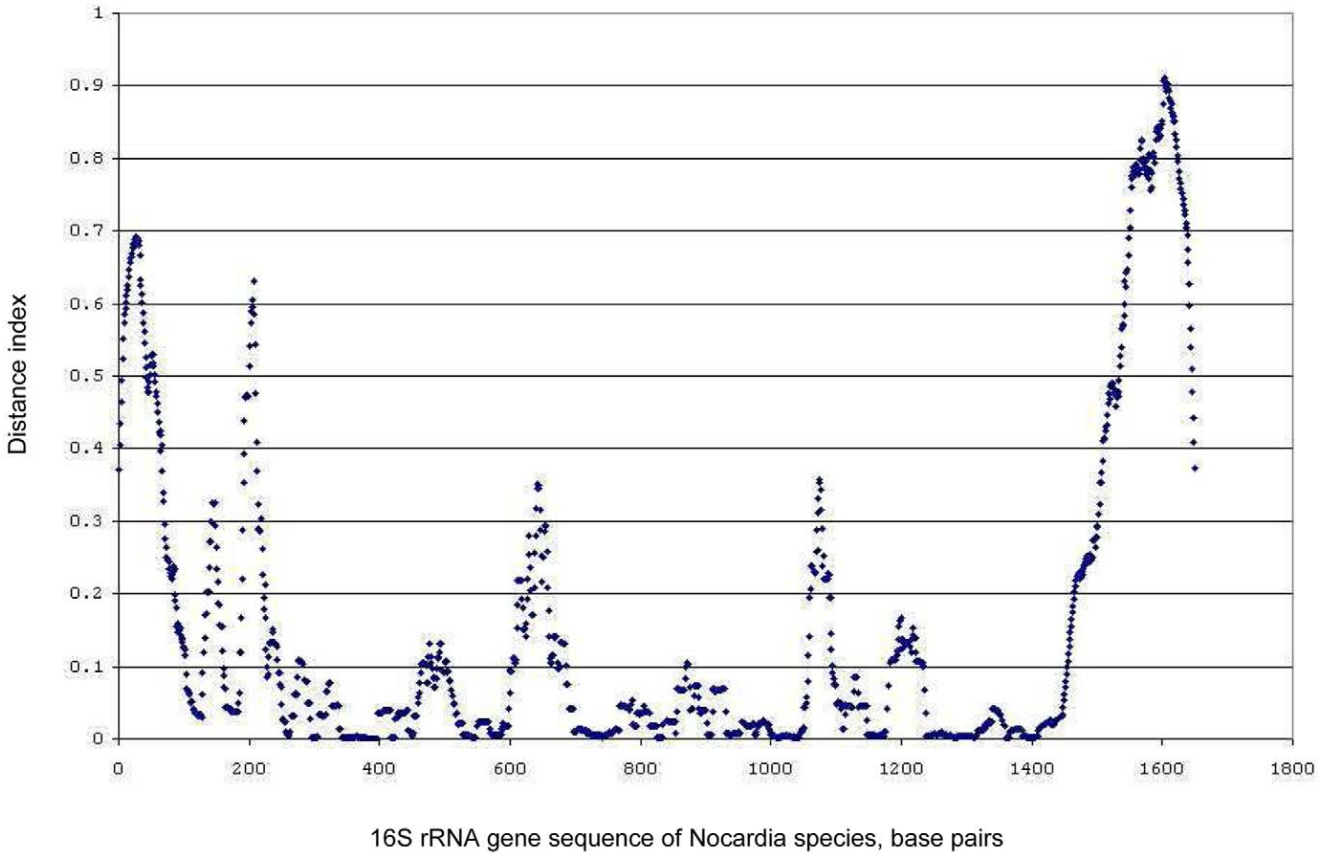

16S rRNA gene sequence of Nocardia species, base pairs

**Figure 2. Relative variability of different regions of the 16S rRNA gene sequence of *Nocardia* species.** Distance matrix generated with a sliding window of 200 nt. Peaks indicate highly variable regions of the gene.

**Figure 3. Principal component analysis (PCA) clustering using the Linear Mapping algorithm over the Euclidian distance of the PCA 1 (x-axis) and PCA 2 (y-axis) scores.** (**A**) and PCA 2 (x-axis) and PCA 3 (y-axis) scores (**B**).
doi:10.1371/journal.pone.0019517.g003

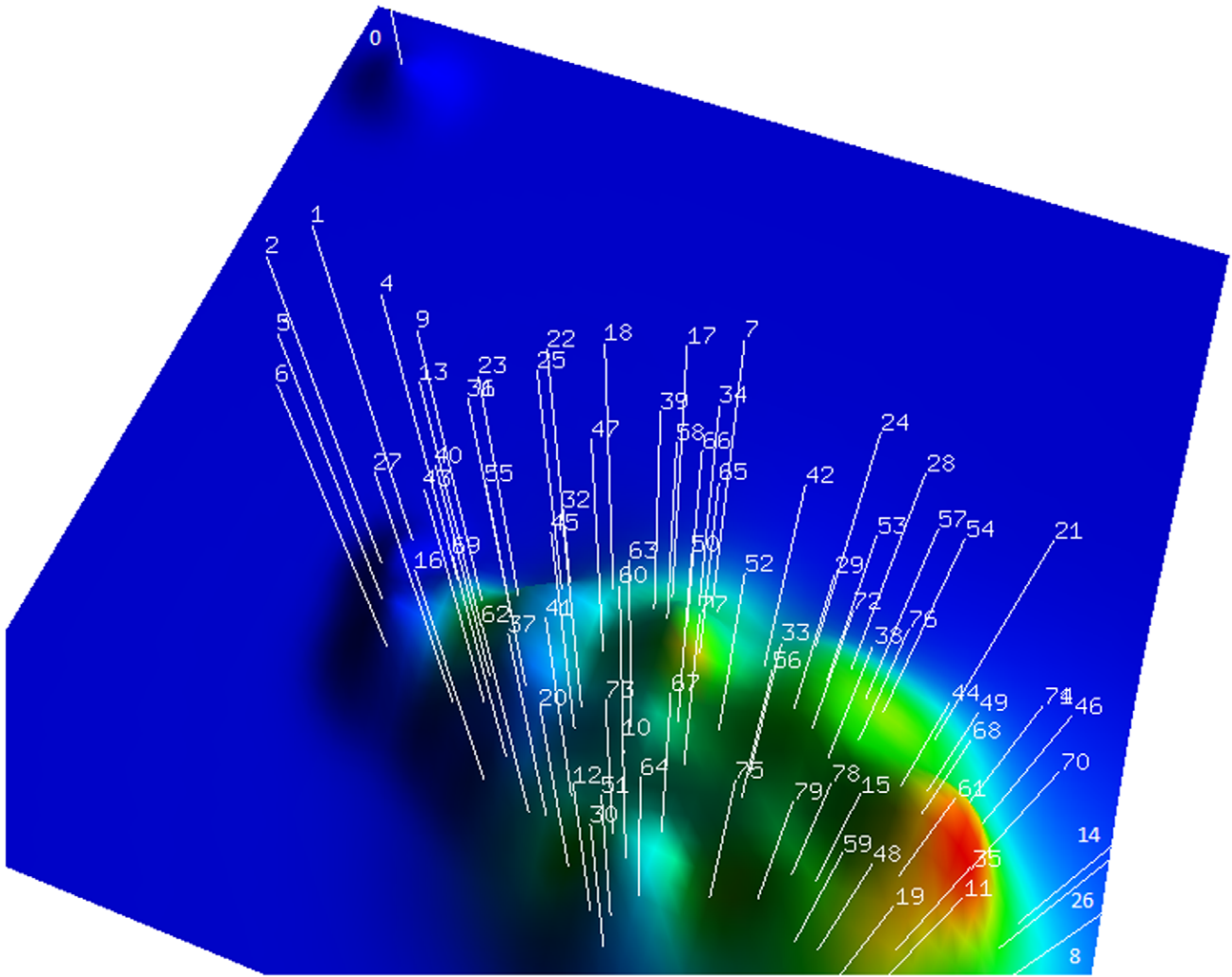

**Figure 4. 'Mountain' view of *Nocardia* species classification where clusters are represented as peaks on the 3D terrain, with the cluster number (starting from 0) pointing to the corresponding mountain peak.** The shape of each peak is a Gaussian curve, which is a rough estimate of the distribution of the data within each cluster. The volume of a peak is proportional to the number of strains contained within the cluster. The height of each peak is proportional to the cluster's internal similarity. The colour of a peak reflects the cluster's internal deviation, where red indicates low deviation where as blue indicates high deviation. Only the colour at the tip of a peak is significant, whereas all other areas colour is determined by blending to create a smooth transition. The numbers indicate the number of a cluster in the experiment (See Supplemental material for details).

doi:10.1371/journal.pone.0019517.g004

**Table 3.** Comparative performance of clustering algorithms.

| Clustering algorithm | Exact match (%)* | Partial match (%)** |
|---|---|---|
| Linear Mapping | 304 (83.52) | 339 (93.13) |
| Cluto | 304 (83.52) | 332 (91.21) |
| Hierarchical Clustering | 294 (80.77) | 320 (87.91) |
| *k*-means | 258+/−0.091 (70.88) | 309 (84.89) |
| PCA | 291 (79.95) | 326 (89.56) |

*Confidence Intervals can be calculated only for *k*-means.
**Partial match was defined as assignment of a sequence to a cluster which contains sequences of *Nocardia* species that match this sequence along with other species.
doi:10.1371/journal.pone.0019517.t003

silhouette values for sets of 27, 36 and 77 clusters were 0.509, 0.460, and 0.438, respectively.

The LM clustering identified three *Nocardia asteroides* clusters ranging from 9 sequences per cluster down to two sequences per cluster. The method produced five clusters for *N. transvalensis*, four clusters for *N. seriolae*, three clusters for *N. cyriacigeorgica* and *N. otitidiscaviarum*, and two clusters for the *N. beijingensis*, *N. concava*, *N. farcinica*, *N. nova*, *N. pseudobrasiliensis*, *N. soli*, and *N. veterana* species. Results for clusters with three or more sequences are shown in Table 1, highlighting the clusters with species name of the reference sequence (or the cluster centroid). One cluster of *N. beijingensis* 16S rRNA gene sequences contained one, probably misclassified, sequence submitted to GenBank in 1996 under the name of *N. asteroides* (Z82228) which was selected as the cluster's centroid (*N. beijingensis* Cluster 2 in Table 1).

A number of clusters contained multiple sequences derived from the same strain, usually a reference strain in a curated culture collection. For example, in a cluster of six 16S rRNA gene sequences of *N. asteroides* only the sequences of three reference strains were represented. Specifically, sequences of different length (accession numbers Z36934, X84850 and DQ659898) were reportedly obtained from ATCC 19247T reference strain and submitted to the GenBank by research groups from the UK in 1994, France in 1995 and the USA in 2006, respectively. Similarly, within this cluster there were two sequences from *N. asteroides* DSM 43757T (X80606 and AF430019) and one sequence from *N. asteroides* GTC 861. 16S rRNA gene sequences in one of the *N. transvalensis* clusters of seven sequences were obtained from four strains (X80609, AF430047 and Z36928 from the strain DSM43405T, X80598 and DQ659916 from ATCC6865T, and Z882232 and Z82249 from N1045 and N630, respectively).

## Identification of reference sequences

The most representative 16S rRNA gene sequences for individual *Nocardia* species that have been identified as 'centroids' in respective clusters, are listed in Table 1. While most of the reference sequences appeared to be the type strains of those species (18 of 27, or 67%), a third (33%) of clusters were assigned with new reference sequences. Of note, certain *Nocardia* species were classified into more than one cluster, and appeared to have more than one reference sequence - for example, for the three *N. asteroides* clusters, the LM clustering method identified the following reference sequences-centroids (Table 1): Z82229 for the largest nine-sequence cluster, and AF163818 and AF430025 for two remaining clusters of two sequences (specified in Supplementary Material).

## Classification of unknown Nocardia 16S rRNA gene sequences

16S rRNA gene sequences with identifications only at the genus level were classified using machine learning algorithms. Two methods were employed from the MATLAB toolboxes (naïve Bayes and *k*-Nearest Neighbour (*k*NN) classifiers) and six methods were used from Weka (naïve Bayes, *k*NN, Boosted Decision Trees, Linear Support Vector Machine (SVM), Cubic SVM, and Quadratic SVM classifiers). Initially, 40 sequences were selected from the centroids of the highly populated clusters, as the validation sequences for the classification algorithms. All methods correctly identified all 40 sequences (see Table S3 for details of the validation). The classification algorithms were then applied to 110 16S rRNA gene sequences of *Nocardia* that had been submitted to the public databases without speciation. The consensus of all classification algorithms was identified, and each method was given a performance score measuring its concordance with the consensus. These scores for the highest scoring methods are shown in Table 4, indicating that the in-house implementation of the k-nearest neighbour method, labelled as simple *k*NN, scored the highest, followed by the MATLAB implementations of naïve Bayes and *k*NN classifiers. The expected frequency of species predicted by the classification algorithms for each of the 110 sequences is also shown in Table 4. This value finds its maximum of 1.0 when all *M* algorithms predicted the same species for each sequence, and its minimum of $1/M$ when every algorithm predicted a different species. Simple *k*NN machine learning classified *Nocardia* species sequences with an accuracy of 92.7% and a mean frequency of 0.578 (Table 4). Simple *k*NN, the MATLAB naïve Bayes and *k*NN classifiers demonstrated the highest frequency of prediction to be greater than 0.6, indicating that the species predicted by these methods will be predicted by almost two-thirds of all methods, averaged over all 110 sequences. In contrast, the lowest expected frequency was for WEKA NBayes and *k*NN at 0.458 and 0.475, respectively, meaning that almost half of the methods agreed on the naming of individual species (Table 4). Even when the classification algorithms predicted different clusters for the same sequence, distances between the centroids of these clusters remained small. Only three sequences were classified to clusters with far-away centroids: AB243007, AF227864, and AJ971864. Unclassified sequences from the GenBank with the results of all classification algorithms and their consensus classification results are listed in Table S4.

## Discussion

We have developed and tested a method for the identification of the most representative sequences from the sets of sequences submitted to public databases. Such sequences may be used as reference sequences to assist the rapid identification in a diagnostic laboratory [4]. This numeric method successfully classified 16S rRNA gene sequences into clusters and identified centroids of the clusters. The clustering of the 364 sequences of 16S rRNA genes of *Nocardia* species revealed the highly homogenous sequence structure of this almost certainly over-classified microbial genus. The high level of similarity of sequences in the *Nocardia* genus explains the relatively narrow range of expected frequencies reached by the majority of machine learning methods. These findings provide a quantitative view of the *Nocardia* genus space with intra- and inter-species variation being captured by similarity and clustering techniques, and confirm the suggestion that genome space is not uniformly filled by a seamless spectrum of intergrading types [22,23]. The recent changes in *Nocardia* taxonomy have provided an indirect validation of our approach and have made our observations even more interesting. For example, a cluster of nine 16S rRNA gene sequences were submitted to the GenBank between 1996 and 2007 by microbiologists from Europe, the USA and Japan under different species names, such as *N. asteroides* (Z82220, Z82221, Z82229, AY191251, DQ659899), *N. transvalensis* (AB084445, AB084446, AB084444) and *N. wallacei* (EU099357). Our algorithm has selected the sequence DQ659899 at the 'centroid' of this cluster. Significantly, the GenBank submission from 2006 was recently changed from *N. asteroides* into *N. wallacei* by its authors, as was the GenBank submission AY191251, making this cluster representative of *N. wallacei* 16S rRNA sequences.

**Table 4.** Performance of classification algorithms against the consensus.

| Method | Agreement with consensus, (%) | Expected frequency of predicted species |
|---|---|---|
| Simple *k*-Nearest Neighbour | 92.73 | 0.669 |
| Matlab NBayes | 88.18 | 0.647 |
| Matlab *k*NN | 87.27 | 0.627 |
| Alignment | 77.22 | 0.592 |
| WEKA NBayes | 57.27 | 0.475 |
| WEKA *k*NN | 52.72 | 0.458 |

Note: The first column presents the accuracy of each method compared to the consensus of all methods used (i.e., the percentage of predictions made by each method that agree with the majority prediction by all methods). The second column shows the expected frequency of the classification made by each method (i.e., the mean frequency of the species predicted by each method, taken over all 110 sequences).
doi:10.1371/journal.pone.0019517.t004

There has been a consensus that assignments to species should be made primarily on the basis of overall sequence similarity, although phenotypic difference should also play a role in fine-scale differentiation [23]. Assignment of a new sequence to a given species is usually done using the sequences of type strains (when available). However, type strains are often defined by the first isolate described for a new species. As a result, historical type strains (for a species) as well as type species (for a genus) may be defined from a clone which is somehow atypical within its clade (see for example the successive redescriptions necessary for many genera). We argue that automated classification process which is based on the sequence of centroids can be more efficient and relevant, even if the centroid sequence is not the sequence of the type strain or type species.

In this context, defining consensus reference sequences for target genes has become a priority for diagnostic laboratories [24]. The finding that most of the reference sequences (66%) appeared to be the type strains of those species reconfirms the validity of our approach. Importantly, a third (34%) of clusters are assigned with new reference sequences. Our study offers new insights into ways in which such reference sequences can be identified. Specifically, it suggests that the LM method might assist best with this task as it performed better than the alternative techniques studied and correctly identified the most representative or reference sequences for each cluster. The LM algorithm also suggested the optimal number of clusters based on given sensitivity parameters, while all other methods required a fixed number of clusters or a fixed distance between clusters to be identified as input. Interestingly, deterministic methods like the LM and Cluto outperformed techniques based on randomization (e.g., *k*-means and hierarchical clustering). This is important as the deterministic methods are required for clustering sequences based on the distance measures produced from multiple sequence alignments.

This study contributes to the debate about the measurement of similarity between sequences [25]. The choice of the most optimal methods of alignment and distance measurement remains controversial [25]. Our experience indicates that progressive MSA based on pair-wise alignments is more suitable for large numbers of phylogenetically-related sequences [26] while long sequences from unrelated organisms could be better compared by more computationally expensive pair-wise alignments [27]. Therefore, this study employed Muscle, a progressive MSA method, for its speed and relative accuracy [28]. Muscle was considered the method of choice for our data sets, which consisted of 364 potentially closely related but not very long sequences.

Importantly, our classification algorithms have been able to assign species names to sequences of *Nocardia* species submitted to the GenBank without proper speciation, and have also been able to identify 16S rRNA gene sequences that have been misclassified. *N. asteroides* sequences were correctly clustered in three different clusters (with the majority of the sequences in the cluster to be identified in GenBank as *N. asteroides*). On the other hand, *N. asteroides* (Z82228) appeared among a cluster of sequences identified by GenBank to be *N. beijingensis*.

The recent advances in computational platforms for microbiologists [3,28–31] have encouraged the development of new tools for the comparative genomics of bacteria. The emergence of high-throughput molecular testing has enabled the potentially more discriminatory and informative 'polyphasic' identification and typing of pathogens that relies on several rather than on only one of target genes, some of which may significantly differ in their speed of evolution [32]. The simultaneous assessment of similarity between multiple genes or loci from sets of microbial genomes within the same or different species has become a new requirement for bioinformatics analyses. For example, in a recent study [33] *Nocardia* speciation was achieved through multilocus sequence analysis (MLSA) of gyrase B the β-subunit of DNA topoisomerase (*gyrB*), 16S rRNA (16S), subunit A of SecA preprotein translocase (*secA1*), 65 kDa heat shock protein (*hsp65*), and RNA polymerase (*rpoB*). Interestingly, three-(*gyrB*+16S+*secA1*) locus MLSA was nearly as reliable as five loci MLSA [33]. While approaches developed in this study should be generalizable to other gene targets and indeed other bacterial species, the sampling bias may add artificial gaps that disrupt any perceived seamless spectrum of species and therefore should be acknowledged as a potential limitation of this study. It also remains to be confirmed whether techniques that have been successful on the clonal microorganism like *Nocardia* would work equally well with other microorganisms, especially with bacteria that are more prone to the lateral gene transfer. Care should be also taken when analysing the diversity of gene sequences submitted to DDBJ/EMBL/GenBank prior to 1995, when automated sequencing technologies with higher fidelity first became available. The quality and length of sequences are important consideration since 16S rRNA based identification techniques may often suffer from a lack of phylogenetic information hitherto retrievable from the short 16S rRNA gene sequences [34]. We also tested other sequences clustering tools such as CD-hit [35] and CLANS (CLuster ANalysis of Sequences) [36]. The first produced two clusters separating the unique *N. seriolae* (AB060281 & AB060282, represented by AB060281) from the rest of the sequences represented by *N. cyriacigeorgica* AB094565. This occurred because the Cd-hit employs a high sequence identity threshold. CLANS uses a version of the Fruchterman–Reingold graph layout algorithm to visualize pair-wise sequence similarities in either two-dimensional or three-dimensional space and couldn't cluster the 364 sequences of average length 1442-bp due to the size of the dataset.

Our method requires neither the number of clusters nor the distances between clusters as input, as these would be difficult to derive for a set of unknown sequences. The method uses two sensitivity parameters: the number of hash codes to map to, and the number of codes to include in one cluster before starting the cut-off. Based on the sensitivity parameters, the number of clusters is produced, and after increasing the parameters to the saturation of the numeric differences in the distance matrix, a maximum number of clusters that can be numerically produced from any given dataset is reached. Similarly, the minimum number of clusters is defined. This feature is potentially very useful for the over- and under-classified datasets of highly clonal microorganisms. Another feature of the method is the identification of the sub-matrices of the distance matrix corresponding to the clusters produced. These sub-matrices can point to the sequence that is the closest to all other sequences in the same cluster to be selected as a centroid.

Two potential limitations of the study should be acknowledged. First, our approach focuses in vertical gene transfer while ignoring the potential impact of lateral gene transfer (LGT). However, the conserved nature of the 16S rRNA gene sequence makes it an unlikely candidate for LGT. Furthermore, some recent evidence indicates the low impact of LGT on 16S rRNA genes (often multi-copy genes with e-values for sequences <10e-80) in contrast to protein-coding genes in a large selection of bacterial genomes [5]. Second, some of the clusters identified in the study had several sequences derived from the same strains and submitted (sometimes by different groups of investigators) to the GenBank over the last two decades. As these GenBank submissions range widely in time, sequence length and potential quality, one would argue that our unsupervised clustering approach should be sufficiently robust to resist the data noise that multiple submissions from the same reference genomes might occasionally introduce. Our study has identified several sequences which were likely misclassified at the time of their submission or which classification had not kept up with changes in *Nocardia* taxonomy. Our study is not a descriptive account on the changing *Nocardia* taxonomy but a proof-of-concept experiment involving mathematically explicit study of sequence similarities with a machine learning component. While filtering for multiple sequences from the same original strains or potentially misclassified sequences has some appeal, this procedure can lead to considerable selection bias.

In conclusion, the identification of the centroids of 16S rRNA gene sequence clusters using novel distance matrix clustering enables the identification of the most representative sequences for each individual species of *Nocardia* and allows the quantitation of inter- and intra-species variability. The LM mapping algorithms based on multiple sequence alignments have demonstrated the important capacity to classify 16S rRNA gene sequences according to microbial species. Our findings have opened new opportunities in the optimisation of bacterial sequence-based species identification by applying sequence alignment and similarity assessments.

## Materials and Methods

### Datasets

The dataset was assembled containing 364 sequences of the 16S rRNA gene of *Nocardia* species deposited in the GenBank [37,38]. The dataset represented 77 different *Nocardia* species whose identification was concordant with the List of Prokaryotic names with Standing in Nomenclature [39]. In addition, 110 16S rRNA gene sequences deposited in DDBJ/EMBL/GenBank but identified only to the *Nocardia* genus level at the time of submission were used for machine learning experiments.

### Clustering of sequences

The distance matrix was calculated for the dataset following the alignment of sequences using the Muscle algorithm [40]. The columns and the rows in this matrix were symmetric, and the distance on the main first diagonal was zero as it matched the same sequence to itself (Figure S1). The second diagonal (connecting the first column with the second row, the second column with the third row, and so forth) represented the distance between a sequence and the most similar sequence.

Four clustering methods were contrasted in the study. First, deterministic clustering was conducted using Cluto [44] with the partitional default functions, as deterministic methods are more relevant in the context of a dataset without missing values and allow the optimisation of clustering functions. The mountain view (Figure 4) produced by the Cluto [44] package was generated by applying Multidimensional Scaling to each of the cluster mid-points, preserving the distances between vertices as they were mapped from a high dimensional space down to a lower dimensional space. Second and third, commonly utilised non-deterministic hierarchical and k-means clustering methods [42] were implemented in MATLAB® [43]. Last, a novel in-house algorithm [45] was compared with the above methods. The latter was based on the linear mapping of the distance matrix diagonal values. Whenever sequences were represented as a 'heat map' with rectangular shapes. The shapes' colours reflected the distance between sequences. The darkest blue shades were congregated along the diagonal as the boundaries around which the natural selection of a cluster should be identified (Figure S1). The process of identifying these boundaries was based on the linear mapping of the second diagonal values to a normalized index value. These index values were then used to detect the break points that separated the clusters.

Clustering trees were constructed using distance measures from multiple sequence alignments [41]. Principal Components Analysis (PCA) was used for the dimensionality reduction of the clustering structure [46]. Two-dimensional plots were constructed for the highest scoring principal components.

### Identification of Reference Sequences

The most representative sequence for each cluster or reference sequence was defined as a 'centroid' of a cluster, from which the distances to all other points (sequences) were minimized. In other words, the reference sequences were ones that were the closest to all other sequences within the same cluster. This was done by identifying the sub-matrices that encapsulated each cluster sequence's pair-wise distances, calculating the total distances from each sequence to all other sequences in the cluster, and identifying the sequence with the minimum total distance as the reference sequence.

### Classification Methods

Naïve Bayes classification and k-nearest neighbour (kNN) algorithms were implemented in MATLAB® (www.mathworks. com). Methods from the Waikato Environment for Knowledge Analysis (WEKA) version 3.6.2 [46] were used to explore predictive models, including boosted decision trees (J48), support vector machines (SVM) with linear, quadratic and cubic kernels, naïve Bayes (NB) classification and kNN. The NB classifiers that assume Gaussian distributions and utilise the 'diaglinear' discriminant function (similar to the "linear" discriminant function, but with a diagonal covariance matrix estimate) were applied. kNN utilised a default Euclidean distance measure between the variables. The distance matrix was employed as a training dataset, considering each column (distances from the column corresponding sequence to all rows' sequences in the dataset) as one variable (parameter). Then, all 16S rRNA sequences from fully identified isolates and the sequences from isolates identified only to a genus level were aligned in one dataset and another distance matrix was generated. This new distance matrix was reorganized so that only the unknown sequences remained in the rows, and only the known sequences remained in the columns. The new 364×110 matrix captured the distances between the unknown sequences and the known sequences, and was used as the testing dataset.

### Method validation and statistical analyses

The random sampling method was used in the K-means algorithm in MATLAB®. The accuracy of the clustering was estimated by the mean value of the silhouette values for each cluster as follows:

$$S(i) = (\min(b(i,:),2) - a(i))./\max(a(i),\min(b(i,:),2)),$$

where $a(i)$ was the average distance from the $i$th point to the other points in its cluster, and $b(i,k)$ was the average distance from the $i$th point to points in another cluster $k$. This value measured the similarity of sequences in one cluster compared to sequences in other clusters, and ranged from −1 to +1. The generalisation performance of machine learning classifiers was evaluated by using 10-fold cross-validation in the testing [47].

## Supporting Information

**Figure S1 Heat map of the *Nocardia* distance matrix based on the Muscle alignment, showing rectangles around potential clusters identified by the Linear Mapping with different sensitivity parameters.**
(TIF)

**Figure S2 Silhouette values for 77 clusters defined by k-means clustering.**
(TIF)

**Table S1 Total numbers of clusters generated by the linear mapping algorithms using different parameters.**
(DOC)

**Table S2 Linear mapping clustering using different colour map sizes and up to 4 colours per cluster for each map size.**
(DOC)

**Table S3 *Nocardia* clustering results for k-means, hierarchical clustering algorithms, Cluto software, and PCA scores, clustered on the Euclidean distance for the first two principal components using the linear mapping clustering algorithm.**
(DOC)

**Table S4 Classification of *Nocardia* species using different machine learning methods.**
(DOC)

## Author Contributions

## References

1. Savolainen V, Cowan RS, Vogler AP, Roderick GK, Lane R (2005) Towards writing the encyclopedia of life: an introduction to DNA barcoding. Philos Trans R Soc Lond B Biol Sci 360: 1805–1811.
2. Hebert PD, Gregory TR (2005) The promise of DNA barcoding for taxonomy. Syst Biol 54: 852–859.
3. Conville PS, Murray PR, Zelazny AM (2010) Evaluation of the Integrated Database Network System (IDNS) SmartGene software for analysis of 16S rRNA gene sequences for identification of *Nocardia* species. J Clin Microbiol 48(8): 2995–2998.
4. Nelson DM, Cann IKO, Altermann E, Mackie RI (2010) Phylogenetic evidence for lateral gene transfer in the intestine of marine iguanas. PLoS ONE 5(5): e10785.
5. Kong F, Chen SCA, Chen X, Sintchenko V, Halliday C, et al. (2009) Assignment of reference 5′-end 16S rDNA sequences and species-specific sequence polymorphisms improves species identification of *Nocardia*. Open Microbiol J 3: 97–105.
6. Grenfell BT, Pybus OG, Gog JR, Wood JLN, Daly JM, et al. (2004) Unifying the epidemiological and evolutionary dynamics of pathogens. Science 303: 327–332.
7. Pybus OG, Rambaut A (2009) Evolutionary analysis of the dynamics of viral infectious disease. Nat Genetics Rev 10: 540–50.
8. Lancashire L, Schmid O, Shah H, Ball G (2005) Classification of bacterial species from proteomic data using combinatorial approaches incorporating artificial neural networks, cluster analysis and principal component analysis. Bioinform 21(10): 2191–2199.
9. Agius P, Kreiswirth BN, Naidich S, Bennett KP (2007) Typing *Staphylococcus aureus* using the *spa* gene and novel distance measures. IEEE Trans Comput Biol Bioinform 4(4): 693–704.
10. Dutilh BE, He Y, Hekkelman ML, Huynen MA (2008) Signature, a web server for taxonomic characterization of sequence samples using signature genes. Nucl Acids Res 36: W470–474.
11. Sgourakis NG, Bagos PG, Papasaikas PK, Hamodrakas SJ (2005) A method for the prediction of GPCRs coupling specificity to G-proteins using refined profile hidden Markov models. BMC Bioinform 6: 104.
12. Karchin R, Karplus K, Haussler D (2002) Classifying g-protein coupled receptors with support vector machines. Bioinformatics 18: 147–159.
13. Baldi P, Chauvin Y, Hunkapiller T, McClure MA (1994) Hidden Markov models of biological primary sequence information. Proc Natl Acad Sci USA 91: 1059–1063.
14. Krogh A, Brown M, Mian I, Sjölander K, Haussler D (1994) Hidden Markov models in computational biology. J Mol Biol 235: 1501–1531.
15. Saubolle MA, Sussland D (2003) Nocardiosis: review of clinical and laboratory experience. J Clin Microbiol 41: 4497–501.
16. Brown-Elliott BA, Brown JM, Conville PS, Wallace RJ, Jr. (2006) Clinical and laboratory features of the *Nocardia* spp. based on current molecular taxonomy. Clin Microbiol Rev 19: 259–82.
17. Roth A, Andrees S, Kroppenstedt RM, Harmsen D, Mauch H (2003) Phylogeny of the genus *Nocardia* based on reassessed 16S rRNA gene sequences reveals underspeciation and division of strains classified as *Nocardia asteroides* into three established species and two unnamed taxons. J Clin Microbiol 41: 851–856.
18. Conville PS, Brown JM, Steigerwalt AG, Brown-Elliott BA, Witebsky FG (2008) *Nocardia wallacei* sp. nov. and *Nocardia blacklockiae* sp. nov., human pathogens and members of the "*Nocardia transvalensis* Complex". J Clin Microbiol 46(4): 1178–1184.
19. Clarridge JE, III (2004) Impact of 16S rRNA gene sequence analysis for identification of bacteria on clinical microbiology and infectious diseases. Clin Microbiol Rev 17: 840–62.
20. Janda JM, Abbott SL (2007) 16S rRNA gene sequencing for bacterial identification in the diagnostic laboratory: pluses, perils, and pitfalls. J Clin Microbiol 45: 2761–2764.
21. Conville PS, Witebsky FG (2007) Analysis of multiple differing copies of the 16S rRNA gene in five clinical isolates and three type strains of *Nocardia* species and implications for species assignment. J Clin Microbiol 45: 1146–1151.
22. Konstantinidis KT, Tiedje JM (2005) Genomics insights that advance the species definition for prokaryotes. Proc Natl Acad Sci USA 102(7): 2567–2572.
23. Doolittle WF, Zhaxybayeva O (2009) On the origin of prokaryotic species. Genome Res 19(5): 744–756.
24. Sintchenko V, Iredell JR, Gilbert GL (2007) Genomic profiling of pathogens for disease management and surveillance. Nat Microbiol Rev 5: 464–470.
25. Edgar R (2010) Big alignments — do they make sense? Robert Edgar's Blog (Available: http://robertedgar.wordpress.com/2010/05/02/big-alignments-do-they-make-sense/).
26. Needleman SB, Wunsch CD (1970) A general method applicable to the search for similarities in the amino acid sequence of two proteins. Mol Biol 48(3): 443–53.
27. Helal M, El-Gindy H, Gaeta G, Sintchenko V (2008) High performance multiple sequence alignment algorithms for comparison of microbial genomes. Proc 19th Int Conf Genome Informatics (GIW 2008), Gold Coast, Australia, 2008.
28. Steinke D, Vences M, Salzburger W, Meyer A (2005) TaxI: a software tool for DNA barcoding using distance methods. Phil Trans Royal Soc B 360: 1975–1980.
29. Davidsen T, Beck E, Ganapathy A, Montgomery R, Zafar N, et al. (2010) The comprehensive microbial resource. Nucleic Acids Res 38: D340–345.
30. Dehal PS, Joachimiak MP, Price MN, Bates JT, Baumohl JK, et al. (2010) MicrobesOnline: an integrated portal for comparative and functional genomics. Nucleic Acids Res 38: D396–400.
31. Markowitz VM, Chen I-MA, Palaniappan K, Chu K, Szeto E, et al. (2010) The integrated microbial genomes system: an expanding comparative anlysis resource. Nucleic Acids Res 38: D382–390.
32. Christen R (2008) Identifications of pathogens—a bioinformatic point of view. Current Opin Biotechnol 19: 266–273.
33. McTaggart LR, Richardson SE, Witkowska M, Zhang SX (2010) Phylogeny and identification of *Nocardia* species based on multilocus sequence analysis. J Clin Microbiol 48(12): 4525–4533.
34. Höfle MG, Flavier S, Christen R, Bötel J, Labrenz M, et al. (2005) Retrieval of nearly complete 16S rRNA gene sequences from environmental DNA following 16S rRNA-based community fingerprinting. Environ Microbiol 2005; 7(5): 670–675.
35. Li W, Godzik A (2006) Cd-hit: a fast program for clustering and comparing large sets of protein or nucleotide sequences". Bioinformatics 22: 1658–1659.
36. Frickey T, Lupas A (2004) CLANS: a Java application for visualizing protein families based on pairwise similarity. Bioinformatics 20(18): 3702–3704.
37. Benson DA, Karsch-Mizrachi I, Lipman DJ, Ostell J, Sayers EW (2010) GenBank. Nucl Acids Res 38(Suppl1): D46–51.
38. Xiao M, Kong F, Sorrell TC, Cao Y, Lee OC, et al. (2010) Identification of pathogenic *Nocardia* species by reverse line blot hybridization targeting the 16S rDNA and 16S–23S rDNA spacer regions. J Clin Microbiol 48(2): 503–511.
39. Euzeby JP (1997) List of bacterial names with standing in nomenclature: a folder available on the Internet (http://www.bacterio.cict.fr/). Int J Syst Bacteriol 47: 590–592.
40. Edgar RC (2004) MUSCLE: multiple sequence alignment with high accuracy and high throughput. Nucl Acids Res 32(5): 1792–1797.

41. Felsenstein J (1989) PHYLIP - Phylogeny Inference Package (Version 3.2). Cladistics 5: 164–166.
42. Fielding AH (2007) Cluster and Classification Techniques for the Biosciences, Cambridge University Press.
43. Gilat A (2004) MATLAB: An introduction with applications. 2nd Edition John Wiley & Sons.
44. Zhao Y, Karypis G (2002) Evaluation of hierarchical clustering algorithms for document datasets. Conf Information & Knowledge Manag, Nov 4–9 2002, McLean, Virginia, USA.

45. Manal H, Kong F, Chen SCA, Zhou F, Dwyer DE, et al. (2011) Linear normalised hash function for clustering gene sequences and identifying reference sequences from multiple sequence alignments. (Submitted).
46. Witten IH, Frank E (2005) Data mining: Practical machine learning tools and techniques, 2nd Edition Morgan Kaufmann, San Francisco.
47. Yeung KY, Ruzzo WL (2001) Principal component analysis for clustering gene expression data. Bioinformatics 17(9): 763–774.